\documentclass[aps,amsfonts,amsmath,prl,tightenlines,twocolumn,showpacs]{revtex4}
\usepackage{epsf}

\def\be{\begin{equation}}
\def\ee{\end{equation}}
\def\beq{\begin{equation}}
\def\eeq{\end{equation}}
\def\bea{\begin{eqnarray}}
\def\eea{\end{eqnarray}}
\def\bml{\begin{subequations}}
\def\blea{\bml\begin{eqnarray}}
\def\elea{\end{eqnarray}\end{subequations}}

\begin{document}

\title{Eternal observers and bubble abundances in the landscape}

\author{Vitaly Vanchurin}
\email{vitaly@cosmos.phy.tufts.edu}
\author{Alexander Vilenkin}
\email{vilenkin@cosmos.phy.tufts.edu}
\affiliation{Institute of
Cosmology, Department of Physics and Astronomy, Tufts University,
Medford, MA  02155}

\begin{abstract}

We study a class of ``landscape'' models in which all vacua have
positive energy density, so that inflation never ends and bubbles of
different vacua are endlessly ``recycled''. In such models, each
geodesic observer passes through an infinite sequence of bubbles,
visiting all possible kinds of vacua. The bubble abundance $p_j$ can
then be defined as the frequency at which bubbles of type $j$ are
visited along the worldline of an observer. We compare this definition
with the recently proposed general prescription for $p_j$ and show
that they give identical results.

\end{abstract}

\pacs{98.80.Cq	
    }

\maketitle 

\section{Introduction}

Nearly all models of inflation are eternal to the future. Once
inflation has started, it continues forever, producing an unlimited
number of pocket universes \cite{Guth,AV83,Linde86}. If there is a
number of different types of pockets, as in the landscape picture
suggested by string theory \cite{Bousso,Susskind}, all the possible
types are produced in the course of eternal inflation. A natural
question is, then, What is the relative abundance $p_j$ of pockets of
type $j$?

This question has proved to be surprisingly difficult to answer. The
total number of pockets is divergent, so one needs to introduce some
sort of a cutoff. If we cut off the count at a constant-time
hypersurface $\Sigma: t={\rm const}$, the resulting abundances are
very sensitive to the choice of the time coordinate $t$ \cite{LM}. The
reason is that the number of pockets in an eternally inflating
universe is growing exponentially with time, so at any time a
substantial fraction of pockets have just nucleated. Which of these
pockets are crossed by the surface $\Sigma$ depends on how the surface
is drawn; hence the gauge-dependence of the result.

A new prescriptions for the calculation of $p_j$, which does not
suffer from the gauge-dependence problem, has been recently suggested
in \cite{GSPVW}. To simplify the discussion, we shall focus on models
where transitions between different vacua occur through bubble
nucleation, so the role of pocket universes is played by bubbles.  To
determine the bubble abundance, one starts with a congruence of
geodesics emanating from some (finite) initial spacelike hypersurface
$\Sigma_0$. As they extend to the future, the geodesics will generally
cross a number of bubbles before ending up in one of the terminal
bubbles, having negative or zero vacuum energy density, where
inflation comes to an end. The geodesics provide a mapping of all
bubbles encountered by the congruence back on the initial
hypersurface.  The proposal is to count only bubbles greater than a
certain comoving size $\epsilon$, and then take the limit $\epsilon\to
0$: 
\be 
p_j \propto \lim_{\epsilon \to 0} N_j(\epsilon).
\label{epsilon}
\ee
Here, $N_j(\epsilon)$ is the number of bubbles of type $j$ with
comoving size grater than $\epsilon$. The comoving size of a bubble is
defined as the size of its image on $\Sigma_0$.

In this prescription, the bubble count is dominated by bubbles formed
at very late times and having very small comoving sizes. (The
asymptotic number of bubbles is infinite even though the initial
hypersurface $\Sigma_0$ is assumed to be finite.) The resulting values
of $p_j$ are independent of the choice of the initial hypersurface,
because of the universal asymptotic behavior of eternal
inflation \cite{foot2}.

An alternative prescription for $p_j$ has been suggested by Easther,
Lim and Martin \cite{ELM}. They randomly select a large number $N$ of
worldlines out of a congruence of geodesics and define $p_j$ as being
proportional to the number of bubbles of type $j$ intersected by at
least one of these worldlines in the limit $N\to\infty$. As the number
of worldlines is increased, the average comoving distance $\epsilon$
between them (on $\Sigma_0$) gets smaller, so most bubbles of comoving
size larger than $\epsilon$ are counted. In the limit of $N\to\infty$,
we have $\epsilon\to 0$, and it can be shown \cite{GSPVW} that this
definition is equivalent to that of \cite{GSPVW} (except in a special
case indicated below). We shall not distinguish between the two
definitions in what follows.

The prescription of \cite{GSPVW,ELM} for $p_j$ has some very
attractive features.  Unlike the earlier prescriptions, it is
applicable in the most general case and does not depend on any
arbitrary choices, such as the choice of gauge or of a spacelike
hypersurface. It is also independent of the initial conditions at the
onset of inflation. It is not clear, however, how uniquely the new
prescription is selected by these requirements. Are there any
alternative prescriptions with the same properties?

In this paper we shall analyze an attractive alternative, which
suggests itself in models including only recyclable (non-terminal)
vacua. In such models each geodesic observer passes through an
infinite sequence of bubbles, visiting all possible kinds of
vacua. The bubble abundance $p_j$ can then be defined as the frequency
at which $j$-type bubbles are visited along the worldline of a given
observer:
\be
	p_j \propto \lim_{\tau \to \infty} N_j(\tau),
\label{tau}
\ee
where $N_j(\tau)$ is the number of times the observer had visited
vacuum $j$ by the time $\tau$. This definition is clearly independent
of gauge or initial conditions. An added attraction here is that the
bubble abundances are defined in terms of observations accessible to a
single observer -- a property that some string theorists find
desirable \cite{Susskind,Bousso2}.

We note that the proposal of Easther, Lim and Martin \cite{ELM} cannot
be applied to models with full recycling \cite{thanks}.  The reason is
that in this case each geodesic worldline intersects an infinite
number of bubbles. We shall therefore focus on the prescription of
Ref. \cite{GSPVW} in what follows.

In the following sections we use the formalism developed in
\cite{GSPVW} to compare the bubble abundances measured by an ``eternal
observer'' with those obtained using the prescription of
\cite{GSPVW}. We find that the two methods give identical results.

\section{Bubble abundances according to \cite{GSPVW}}

In this section we closely follow the analysis given in \cite{GSPVW},
specializing it to the case of fully recyclable vacua.

The fraction of comoving volume $f_j(t)$ occupied by vacuum of type
$j$ at time $t$ is given by the evolution equation \cite{recycling},
\be
	 \frac{d f_i(t)}{dt}=\sum_{j=1}^n M_{ij}f_j,
\label{evolution}
\ee
where 
\be
	M_{ij}=\kappa_{ij}-\delta_{ij}\sum_{r=1}^n \kappa_{ri},
\label{M}
\ee and $\kappa_{ij}$ is the probability per unit time for an observer
who is currently in vacuum $j$ to find herself in vacuum
$i$.  $f_i$ are assumed to be normalized as 
\be 
\sum_{i=1}^n f_i = 1.  
\ee

The magnitude of $\kappa_{ij}$ depends on the choice of the time
variable $t$ \cite{recycling}. The most convenient choice for our
purposes is to use the logarithm of the scale factor as the time
variable; this is the so-called scale-factor time,
\be
a(t)\equiv e^t.
\label{time}
\ee
With this choice \cite{foot1},
\be
\kappa_{ij}=(4\pi/3)H_j^{-4}\Gamma_{ij},
\label{kappa}
\ee
where 
\be
\Gamma_{ij}=A_{ij}e^{-I_{ij}-S_j},
\ee
$I_{ij}$ is the tunneling instanton action, 
\be
S_j=\frac{\pi}{H_j^2}
\ee
is the Gibbons-Hawking entropy of $j$-th vacuum, and $H_j$ is the
corresponding expansion rate. The instanton action and the prefactor
$A_{ij}$ are symmetric with respect to interchange of $i$ and $j$
\cite{LW}. Hence, we can write
\be
\kappa_{ij}=\lambda_{ij}H_j^{-4}e^{-S_j}
\label{kappaij}
\ee
with
\be
\lambda_{ij}=\lambda_{ji}.
\label{lambda}
\ee

Assuming that all vacua are recyclable and that the matrix
$M_{ij}$ is irreducible (each vacuum is accessible from every
other one), it can be shown \cite{recycling,GSPVW} that
Eq.(\ref{evolution}) has a unique stationary solution with $df_j/dt
=0$ and 
\be
\sum_{j=1}^n M_{ij}f_j =0.
\label{Ms}
\ee
In fact, the solution can be found explicitely:
\be
f_j\propto H_j^4 e^{S_j}.
\label{solution}
\ee 
This can be easily verified by substituting (\ref{solution}) in
(\ref{Ms}),(\ref{M}) and making use of (\ref{kappaij}) and
(\ref{lambda}).

$f_j$ has the meaning of the fraction of time spent by a geodesic
observer in bubbles of type $j$. As one might have expected,
Eq.(\ref{solution}) shows that it is proportional to the statistical
weight of the corresponding vacuum, $\exp(S_j)$.

We shall now use the prescription of Ref. \cite{GSPVW} to determine
the bubble abundance.  The increase in the number of $j$-type bubbles
due to jumps from other vacuum states in an infinitesimal time
interval $dt$ can be expressed as
\be
	dN_j(t) = \sum_{i=1}^n \frac{\kappa_{ji} f_i}{\frac{4
	\pi}{3}R_i(t)^3}dt.
\label{increase}
\ee
Here, $R_i(t)$ is the comoving radius of the bubbles nucleating in
vacuum $i$, which is set by the comoving horizon size at the time $t$
of bubble nucleation,
\be
R_i(t)=H_i^{-1}a^{-1}(t)= H_i^{-1}e^{-t},
\ee
where $a(t)$ is the scale factor and we have used the definition of
scale-factor time in (\ref{time}).

Bubbles of comoving size greater than $\epsilon$ are created at
$t<-ln(\epsilon H_i)$. Integrating Eq.(\ref{increase}) up to this
time, we obtain 
\be
N_j = \frac{ \epsilon^{-3}}{4 \pi}\sum_{i=1}^n \kappa_{ji} f_i.
\ee
The prescription of \cite{GSPVW} is that $p_j\propto N_j$,
and thus
\be
	p_j \propto \sum_{i=1}^n \kappa_{ji} f_i \propto 
\sum_{i=1}^n \lambda_{ji},
\label{p1}
\ee
where we have used Eqs. (\ref{solution}) and (\ref{kappaij}).

\section{Eternal observers}

We consider a large ensemble of eternal observers.
They evolve independently of one another, yet statistically all of
them are equivalent.  The worldline of each observer can be
parametrized by discrete jumps to different vacuum states, so the time
variable $\tau$ takes values in natural numbers, $\tau =
1,2,3...$, and is incremented by one whenever the observer jumps to
a different vacuum state.

Let $x_j(\tau)$ be the fraction of observers in vacuum $j$ at ``time''
$\tau$. $x_j(\tau)$ is normalized as
\be
        \sum_{j=1}^n x_j = 1
\ee
and satisfies the evolution equation
\be
x_i(\tau+1)=\sum_{j=1}^n T_{ij}x_j(\tau),
\label{eternal}
\ee
where the transition matrix is given by 
\be
	T_{ij} = \frac{\kappa_{ij}}{\kappa_j}
\ee
and
\be
\kappa_j = \sum_{r=1}^n \kappa_{rj}.
\ee
The diagonal elements of the transition matrix are exactly zero,
\be
	T_{ii}=\kappa_{ii}=0, 
\ee
since we require each observer to jump to some other vacuum at every
time step.

In the case of complete recycling that we are considering here, one
expects that the evolution equation (\ref{eternal}) has a
stationary solution satisfying
\be
\sum_{j=1}^n (T_{ij}-\delta_{ij})x_j =0.
\label{stationary}
\ee
And indeed, rewriting Eq.(\ref{stationary}) as
\be
\sum_{j=1}^n M_{ij}(x_j/\kappa_j)=0,
\label{stationary1}
\ee
and comparing with Eq.(\ref{Ms}), we see that the stationary solution
of (\ref{stationary1}) is
\be
x_j = \kappa_j f_j.
\label{xj}
\ee
Here, $f_j$ is the solution of (\ref{Ms}), which is given by
(\ref{solution}). 

Suppose now that we have an ensemble of observers described by the
stationary distribution (\ref{xj}). Since the sequences of vacua
visited by all observers are statistically equivalent, it is not
difficult to see that the distribution of vacua along the observer's
worldlines is given by $p_j \propto x_j$, or
\be
	p_j  \propto  \sum_{i=1}^n \kappa_{ij} f_j 
\label{abundance}
\ee

Using Eq.(\ref{Ms}) with $M_{ij}$ from (\ref{M}), we have
\be
	0 = \sum_{i=1}^n M_{ji} f_i =
	\sum_{i=1}^n \kappa_{ji} f_i - \sum_{i=1}^n \kappa_{ij} f_j, 
\ee
or
\be
	\sum_{i=1}^n \kappa_{ji} f_i = \sum_{i=1}^n \kappa_{ij} f_j. 
\ee
Therefore, Eq.~(\ref{abundance}) can be also rewritten as
\be
	p_j  \propto \sum_{i=1}^n \kappa_{ji} f_i,
\label{worldline}
\ee
which is identical to (\ref{p1}).

\section{Discussion}

In this paper we considered a special but relatively wide class of
models in which all vacua have positive energy density and are
therefore inflationary. Transitions between different vacua occur
through bubble nucleation, and each geodesic worldline encounters an
infinite sequence of bubbles. The bubble abundance can then be defined
as the frequency at which bubbles of a given type are encountered in
this sequence. We have shown that this natural definition is
equivalent (in this class of models) to the prescription of
Ref. \cite{GSPVW} (which has greater generality).

We wish to emphasize the difference between the stationary
distribution $f_j$ [ Eq.(\ref{solution})] and the bubble abundance
$p_j$, which has been the focus of our attention here. The difference
is very striking in the case when there are only two vacua. Then
Eq.(\ref{p1}) gives
\be
p_1/p_2 =\lambda_{12}/\lambda_{21}=1,
\ee
while Eq.(\ref{solution}) gives
\be
f_1/f_2=(H_1/H_2)^4 e^{S_1-S_2}.
\ee
The stationary solution $f_j$ strongly favors lower-energy vacuum,
which has a higher entropy $S_j$, while the distribution $p_j$ seems
to indicate that the two vacua are equally abundant. The prescription
of \cite{GSPVW} was recently criticized in \cite{Bousso2} for
failing to give probabilities proportional to exponential of the
entropy.

We note, however, that the distributions $f_j$ and $p_j$ have very
different meanings. $f_j$ is proportional to the average time a
geodesic observer spends in vacuum $j$ before transiting to another
vacuum. If, as a result of quantum fluctuations, the horizon region
accessible to the observer scans all of its quantum states, spending
roughly equal time in each of them, then one expects $f_j \propto
\exp(S_j)$. This is indeed the case, up to a prefactor. On the other 
hand, $p_j$ is the frequency at which a given vacuum $j=1,2$ appears
in the vacuum sequence along a geodesic worldline. In the case of only
two vacua, the sequence is $1,2,1,2,1,...$, and it is clear that both
vacua occur with the same frequency.

The prescription of \cite{GSPVW} for the bubble abundance is just a
proposal. It was not derived from first principles, and its validity
would be put into question by any alternative proposal satisfying the
necessary invariance and common sense requirements. We therefore find
it reassuring that this prescription turned out to be equivalent to
that of \cite{ELM} and to the ``eternal observer'' proposal in their
respective ranges of validity.

The bubble abundance is necessary for the calculation of probabilities
of various measurements in the landscape. The full expression for the
probability includes the volume expansion factor inside the bubbles
and the density of observers, in addition to $p_j$. For a detailed
discussion of these factors, see \cite{GSPVW,AV06}.

\section*{Acknowledgments}

We are grateful to Jaume Garriga and Serge Winitzki for useful
comments and discussions.  This work was supported in part by the
National Science Foundation.


\begin{thebibliography}{9}
\bibitem{foot2}
We assume that any vacuum is accessible through bubble nucleation from
any other (non-terminal) vacuum. Alternatively, if the landscape
splits into several disconnected domains which cannot be accessed from
one another, each domain will be characterized by an independent
probability distribution, and our analysis will still be applicable to
any of these domains.
\bibitem{foot1} 
Throughout the paper we use Planck units. Einstein's summation
convention is not used: all summations are explicitely indicated.
\end{thebibliography}
\end{document}